\documentclass[11pt,fleqn]{article}

\usepackage{amsfonts}
\usepackage{amsmath}
\usepackage{amssymb}
\usepackage{mathtools}
\usepackage{mathrsfs}
\usepackage{enumerate}
\usepackage{bbm}
\usepackage{graphicx,epsfig,amsthm,color}
\usepackage{pstricks}
\usepackage{subfig}
\usepackage{graphicx}
\usepackage{units}

\usepackage{caption}
\usepackage{booktabs}
\usepackage{textcomp}
\usepackage{array}
\usepackage{cite}
\usepackage{cancel}

\date{\today}

\newcolumntype{z}[1]{>{\RaggedRight\hspace{0pt}}p{#1}}
\newcolumntype{w}[1]{>{\RaggedRight\hspace{0pt}}p{#1}}
\newcolumntype{v}[1]{>{\Centering\hspace{0pt}}p{#1}}

\def \la {\lambda}

\def \a {\alpha}

\def\be{\begin{equation}}
\def\ee{\end{equation}}
\def\bea{\begin{eqnarray}}
\def\eea{\end{eqnarray}}

\def\be{\begin{equation}}
\def\ee{\end{equation}}
\def\bea{\begin{eqnarray}}
\def\eea{\end{eqnarray}}

\def\erp2{{\rm e}^{2\rho}}
\def\erm2{{\rm e}^{-2\rho}}
\def\er4{{\rm e}^{4\rho}}

\def\be{\begin{equation}}
\def\ee{\end{equation}}
\def\bea{\begin{eqnarray}}
\def\eea{\end{eqnarray}}

\def\m0{m_{\nu_{0,i}}}

\def\T0{T_{\nu_0}}

\newcommand{\half}{\frac{1}{2}}

\newcommand{\beqa}{\begin{eqnarray}}
\newcommand{\eeqa}{\end{eqnarray}}
\newcommand{\bpr}{\begin{problem}}
\newcommand{\epr}{\end{problem}}
\newcommand{\bcent}{\begin{center}}
\newcommand{\ecent}{\end{center}}
\newcommand{\bfig}{\begin{figure}}
\newcommand{\efig}{\end{figure}}
\newcommand{\bpc}{\begin{picture}}
\newcommand{\epc}{\end{picture}}

\newcommand{\nnb}{\nonumber}
\newcommand{\reflef}{(\ref}

\newcommand{\nn}{\nonumber}

\renewcommand{\and}{A_{0}^{\nu ,D}(s)}

\newcommand{\bee}{\begin{equation}}

\def\beq{\begin{eqnarray}}
\def\eeq{\end{eqnarray}}

\newcommand{\Ga}{\Gamma}

\newcommand{\lmd}{\lambda}
\newcommand{\bright}{\begin{flushright}}
\newcommand{\eright}{\end{flushright}}
\newcommand{\bminip}{\begin{minipage}}
\newcommand{\eminip}{\end{minipage}}

\DeclareCaptionLabelSeparator{mysep}{\hspace{3pt}:\hspace{3pt}}
\DeclareCaptionLabelFormat{mypiccap}{Fig.\hspace{3pt}{#2}}
\DeclareCaptionLabelFormat{mytabcap}{Tab.\hspace{3pt}{#2}}

\begin{document}

\date{}
\title{
\vskip 2cm {\bf\huge Species, chameleonic strings and the concept of particle}\\[0.8cm]}

\author{
{\sc\normalsize Andrea Zanzi\footnote{Email: zanzi@th.physik.uni-bonn.de}\!\!}\\[1cm]
{\normalsize Via de' Pilastri 34, 50121 Firenze -  Italy}\\}
 \maketitle \thispagestyle{empty}
\begin{abstract}
{We recently proposed a chameleonic solution to the cosmological constant problem - Phys. Rev. D82 (2010) 044006. We further elaborate on our proposal discussing also some new results. The basic ingredients of the model are A) a chameleonic string dilaton parametrizing the amount of scale symmetry of the problem in the Einstein frame; B) a dual nature of the concept of particle where a splitting in local and global components is introduced (for every local particle a corresponding global particle is present in the theory). In this paper we proceed stepwise: 1) we point out that the value of the chameleonic dilaton in the Einstein frame parametrizes also the string length and, in particular, the string mass is an increasing function of the matter density; 2) the concept of global particle is clarified (indeed global particles are simply cosmic strings); 3) in the last quantization step, in the Feynman diagrams of our model, the UV cut-off (which is chosen to be the string mass) is a function of the chameleonic dilaton (naturally this stringy regularization is fully compatible with the chameleonic scale invariance of the model); 4) we show that a large number of particle species might be useful to keep locally under control the dangerous variations of the fundamental constants, however, more research efforts are necessary to make this point clear; 5) the chameleonic behaviour of matter fields and of the de Broglie wavelength of local matter particles is pointed out; 6) we briefly touch upon some ideas (which are still waiting for a full confirmation) regarding a potential connection between the shortest length scale of nature and the cosmological constant. A detailed phenomenological analysis of the entire model is required to test these ideas.\\
}
\end{abstract}
%PACS numbers: 04.60.Cf, 98.80.-k, 95.36.+x

\clearpage

\tableofcontents
\newpage

%%%%%%%%%%%%%%%%%%%%%%%%%%%%%%%%%%%%%%%%%%%%%%%%%%%%%%%%%%%%%%%%%%%%%%%%%%%%%%%%%%%%%%%%%%%%%%%%%%%%%%%%%%%%%%%%%%%%%
% Section INTRODUCTION
%%%%%%%%%%%%%%%%%%%%%%%%%%%%%%%%%%%%%%%%%%%%%%%%%%%%%%%%%%%%%%%%%%%%%%%%%%%%%%%%%%%%%%%%%%%%%%%%%%%%%%%%%%%%%%%%%%%%%
\setcounter{equation}{0}
\section{Introduction}

One of the main problems in modern Cosmology is the {\it cosmological constant} one \cite{Weinberg:1988cp} (for a review see
\cite{Nobbenhuis:2006yf}). The observational evidence of the accelerated nature of the cosmic expansion \cite{Riess:1998cb,
Perlmutter:1998np, Spergel:2003cb, deBernardis:2000gy} may be explained by a (properly chosen) constant term in the gravitational action of Einstein's General Relativity (GR). Among the many possibilities already discussed in the literature to account for this acceleration (for a review see for example \cite{Copeland:2006wr, Brax:2009ae}), the presence of the cosmological constant can be considered, on the one hand, the simplest explanation from the observational point of view, on the other hand, extremely challenging from the fundamental point of view. The origin of the problem is rooted in the degeneracy between a constant term in the Einstein's action and a vacuum energy term \cite{Zeldovich:1967gd, Zeldovich:1968zz}. We can therefore define an {\it effective} cosmological constant given by 
\bea 
\Lambda_{eff} = \Lambda + 8 \pi G <\rho> 
\eea 
where $\Lambda$ is the {\it bare} (or classical) cosmological constant, while the $<\rho>$-term is the energy density of the vacuum. In this way the bare cosmological constant is {\it dressed} by quantum corrections, analogously to all other physical parameters in quantum field theory (QFT). If one believes QFT to be correct up to the Planck scale (i.e. $M_p \simeq 10^{19} GeV$), then this scale provides a natural UV cut-off and this would give us a vacuum energy density of $(10^{19} GeV)^4=10^{76} GeV^4$, which is roughly 123 orders of magnitude larger than the currently observed value 
\bea
\rho_{vac}\simeq 10^{-47} GeV^4 \simeq 10^{-29} g/cm^3 \simeq (0.1 mm)^{-4} \simeq (10^{-12} s)^{-4}.
\eeq
This means that the bare cosmological constant needs to be extremely fine-tuned in order to obtain the correct physical result. In conclusion the question is: {\it why the effective cosmological constant is so close to zero?}

In a recent paper \cite{Zanzi:2010rs} we solved this problem from the standpoint of string theory (see also \cite{lettera}). The solution is obtained by mixing together some of the ideas currently known by the physics community to account for the cosmic accelerated expansion. Among them, we mention: 1) a modification of GR at large distances (see for example \cite{Dvali:2000hr}); 2)
backreaction effects \cite{Rasanen:2003fy, Kolb:2004am,
Kolb:2005da}; 3) a dynamic Dark Energy (DE) fluid. Let us start considering the element we
mentioned last. Scalar degrees of freedom are
a common feature in physics beyond the Standard Model (SM), for
example, they can be related to the presence of extra-dimensions. The Dark Energy could be the manifestation of an
ultralight scalar field rolling towards the minimum of its
potential \cite{Ratra:1987rm, Wetterich:1994bg, Zlatev:1998tr,
Carroll:1998zi}. Remarkably, there are reasons to maintain a
non-trivial coupling between the scalar field and matter, for
instance: a) to solve, at least partially, the coincidence
problem, a direct interaction between Dark Matter (DM) and DE has
been discussed \cite{Amendola:1999er, Amendola:2000uh,
TocchiniValentini:2001ty, Comelli:2003cv, Pietroni:2002ey,
Huey:2004qv, Amendola:2004ew, Gasperini:2003tf}; b) string theory
suggests the presence of scalar fields (dilaton and moduli)
coupled to matter (for an introduction see for example
\cite{Becker:2007zj, Gaillard:2007jr}). Consequently, a direct
interaction between matter and an ultralight scalar field can be
welcome. However, this could be phenomenologically dangerous:
violations of the equivalence principle (as far as the dilaton field is concerned the reader is referred to
\cite{Taylor:1988nw, Damour:1994ya, Damour:1994zq, Damour:2002nv, Damour:2002mi, Kaplan:2000hh, Damour:2010rp}), time dependence of couplings (for reviews
see \cite{Uzan:2010pm, Fischbach:1999bc}).

One possible way-out is to consider "chameleon scalar fields"
\cite{Khoury:2003aq, Khoury:2003rn, Mota:2003tc},
namely scalar fields coupled to matter (including the baryonic
one) with gravitational (or even higher) strength and with a mass
dependent on the density of the environment. On cosmological
distances, where the densities are very small, the chameleons are
ultralight and they can roll on cosmological time scales. On the
Earth, on the contrary, the density is much higher and the field
is massive enough to satisfy all current experimental bounds on deviations from GR. 
In other words, the physical properties of this field
vary with the matter density of the environment and, therefore, it
has been called chameleon. The "chameleon mechanism" can be considered as a (local) 
stabilization mechanism which exploits the interaction matter-chameleon. Our solution to the cosmological constant problem discussed in \cite{Zanzi:2010rs, lettera} is obtained through these ideas: the solution is based on the chameleonic behaviour of the string dilaton\footnote{Many other stabilization
mechanisms have been studied for the string dilaton in the
literature. In particular, as far as heterotic string theory is
concerned, we can mention: the racetrack mechanism
\cite{Krasnikov:1987jj, Casas:1990qi}, the inclusion of
non-perturbative corrections to the Kaehler potential
\cite{Casas:1996zi, Binetruy:1996xja, Barreiro:1997rp}, the
inclusion of a downlifting sector \cite{Lowen:2008fm}, Casimir energy \cite{Zanzi:2006xr}.} \cite{Zanzi:2010rs}.

In the string frame (S-frame) of our model of reference \cite{Zanzi:2010rs}, the cosmological constant is very large and the dilaton is stabilized, while, after a conformal transformation to the Einstein frame (E-frame), the
dilaton is a chameleon and it is parametrizing the
amount of scale symmetry of the problem. Therefore, the E-frame cosmological constant is under control for symmetry reasons. For a detailed discussion of the conformal transformation see also \cite{lettera}. This result points out a non-equivalence of different conformal frames at the quantum level (the cosmological constant is under control only in the E-frame). In the literature, scale invariance has
already been analyzed in connection to the cosmological constant problem (see for example \cite{Wetterich:2008sx, Wetterich:2009az, Wetterich:2008bf, Wetterich:2010kd} and references therein). In our
scenario \cite{Zanzi:2010rs}, the chameleonic behaviour of the field implies that
all the
usual contributions to the vacuum energy (from supersymmetry
[SUSY] breaking, from axions, from electroweak symmetry
breaking...) are extremely large with respect to the meV-scale
only {\it locally}, while on cosmological distances (in the
E-frame) they are suppressed.

In our model of reference \cite{Zanzi:2010rs}, all the particles p=p(x) (dilaton, gravitons, matter fields,...) are split into a fluctuating and a background component $p=p_{f}(x) + p_{b}(t)$ and the Feynman diagrams which are relevant for the cosmological constant are constructed exploiting only the background component of the fields. In other words, we introduced a dual nature of the concept of particle different from the standard well-known particle-wave duality\footnote{In the concluding remarks of this paper, the reader will find a short summary of a research project where we suggest to investigate the potential identification of the probability waves of quantum mechanics with waves on (quantum) matter strings.}: we introduced \cite{Zanzi:2010rs} a local-global dual nature of particles (i.e. its density dependence). In our proposal, local particles describe the physical phenomena locally, in this room, while global particles are relevant cosmologically. The two aspects of the particles are related to each other and, in particular, effects from the fluctuating to the global particles are useful counterterms in the renormalization process. Another result of reference \cite{Zanzi:2010rs} should be mentioned. Typically, quantum field theory (QFT) calculations of loop Feynman diagrams produce, as a final result, a {\it constant} result. In our approach of reference \cite{Zanzi:2010rs}, on the contrary, the result of a loop calculation is a {\it function} of the length scale (i.e. the local loop is different with respect to the cosmological one) and this is due to the different amount of scale invariance.

At this stage several problems in the model of references \cite{Zanzi:2010rs, lettera} must be faced, for example:
\begin{itemize}
\item the careful reader may be worried about a chameleonic behaviour of the string dilaton, because a shift of $\sigma$ could induce a shift in the fundamental constants. For example, the density gap between the atmosphere and a common solid object could induce an unacceptable shift of the string coupling $g_s$.
\item The new dual nature of the particles (local and global aspects) should be further discussed.
\item In our model we quantize the theory many times: we consider three different quantization steps (see \cite{Zanzi:2010rs}). The density dependence of the amount of scale symmetry of the system in the E-frame guarantees that the evaluation of quantum loops does {\it not} produce a constant result. As already mentioned above, quantum loops are {\it functions} of the length scale in our model. This peculiar result can lead us to different regularization scenarios. One first possibility is to introduce in the model a simple constant cut-off and to imagine that, for some reason, the larger is the matter density, the larger is the degree of divergence of the loops. For example we can imagine a quadratic divergence for a loop constructed with local particles, but a logarithmic divergence for the same loop constructed with global particles. A different possibility is to imagine that the functional dependence of the loops on the cut-off is fixed, but, for some reason, the cut-off is a function of the length scale. It would be rewarding to remove this ambiguity between the two scenarios and to make the regularization mechanism clear and easy to understand. 
\end{itemize}

In this paper, we will proceed stepwise towards a (partial) solution of the problems.\\ 1) As far as the variation of the fundamental couplings is concerned, we will exploit a recent paper by G. Veneziano \cite{Veneziano:2001ah} where a saturation mechanism for the coupling constants has been discussed in the framework of strongly coupled string theory with a large number of species. Here we create a link between the model of reference \cite{Veneziano:2001ah} and the one in \cite{Zanzi:2010rs}. The connection with the Veneziano's model might be useful to keep locally under control the dangerous variation of the fundamental constants.\\ 2) A chameleonic behaviour of the string length is pointed out and, therefore, the dual nature of the concept of particle is easy to understand (local particles are small interacting strings, while global particles are simply cosmic strings and they are almost non-interacting).\\ 3) The chameleonic behaviour of the string mass provides a natural varying cut-off for the quantum Feynman diagrams (of the last quantization step) and it sheds some light on the details of the regularization mechanism (which is fully compatible with the restoration of scale invariance).\\ 4) The chameleonic behaviour of matter particles and of the de Broglie wave-length of local matter particles is pointed out. More research efforts are necessary to clarify whether the phenomenological aspects of this proposal are under control.

About the organization of this paper, in section \ref{CC} we review our model of reference \cite{Zanzi:2010rs, lettera}; in section \ref{openp} we draw some intermediate conclusions and we summarize the problems already mentioned above; in section \ref{resolution} the Veneziano's model \cite{Veneziano:2001ah} is connected to the model of reference \cite{Zanzi:2010rs} and the partial solution to the three problems mentioned above is presented. In section \ref{debroglie}, the chameleonic behaviour of matter particles and of the de Broglie wavelength of local particles is pointed out. In section \ref{mixing} we touch upon a few elements of a research project that we consider particularly interesting. It concerns the potential connection between the shortest length scale of nature and the cosmological constant. Some clues of this connection are already present in our model and we summarize them in section \ref{mixing}. However, more research efforts are necessary to render this connection clear.  Section \ref{conclusions} contains some concluding remarks.

\setcounter{equation}{0}
\section{The model}
\label{CC}

In this section we will briefly summarize our stringy solution to the
cosmological constant problem presented recently in \cite{Zanzi:2010rs}. For more details see \cite{Zanzi:2010rs, lettera}.

\subsection{The action}
\label{modello}

Our starting point is the string-frame, low-energy, gravi-dilaton
effective action, to lowest order in the $\a'$ expansion, but
including dilaton-dependent loop (and non-perturbative)
corrections, encoded in a few  ``form factors" $\psi(\phi)$,
$Z(\phi)$, $\alpha{(\phi)}$, $\dots$, and in an effective dilaton
potential $V(\phi)$ (obtained from non-perturbative effects). In
formulas (see for example \cite{Gasperini:2001pc} and references therein):
\bea S &=& -{M_s^{2}\over 2} \int d^{4}x \sqrt{-  g}~
\left[e^{-\psi(\phi)} R+ Z(\phi) \left(\nabla \phi\right)^2 +
{2\over M_s^{2}} V(\phi)\right]
\nonumber \\
&-& {1\over 16 \pi} \int d^{4}x {\sqrt{-  g}~  \over
\alpha{(\phi)}} F_{{\mu\nu}}^{2} + \Ga_{m} (\phi,  g, \rm{matter})
\label{3} \eea Here $M_s^{-1} = \la_s$ is the fundamental
string-length parameter and $F_{\mu\nu}$ is the gauge field
strength of some fundamental grand unified theory (GUT) group ($\a(\phi)$ is the
corresponding gauge coupling). We imagine having already
compactified the extra dimensions and having frozen the corresponding
moduli at the string scale.

Since the form factors are {\it unknown} in the strong coupling
regime, we are free to {\it assume} that the structure of these
functions in the strong coupling region implies an S-frame
Lagrangian composed of two different parts: 1) a scale-invariant
Lagrangian ${\cal L}_{SI}$. This part of our lagrangian has
already been discussed in the literature by Fujii in references
\cite{Fujii:2002sb, Fujii:2003pa}; 2) a Lagrangian which
explicitly violates scale-invariance ${\cal L}_{SB}$.

In formulas we write:

\beq {\cal L}={\cal L}_{SI} + {\cal L}_{SB}, \label{Ltotale}\eeq where the
scale-invariant Lagrangian is given by:

\begin{equation}
{\cal L}_{\rm SI}=\sqrt{-g}\left( \half \xi\phi^2 R -
    \half\epsilon g^{\mu\nu}\partial_{\mu}\phi\partial_{\nu}\phi -\half g^{\mu\nu}\partial_\mu\Phi \partial_\nu\Phi
    - \frac{1}{4} f \phi^2\Phi^2 - \frac{\lambda_{\Phi}}{4!} \Phi^4
    \right).
\label{bsl1-96}
\end{equation}
$\Phi$ is a scalar field representative of matter fields,
$\epsilon=-1$, $\left( 6+\epsilon\xi^{-1} \right)\equiv
\zeta^{-2}\simeq 1$, $f<0$ and $\lambda_{\Phi}>0$.
One may write also terms like $\phi^3 \Phi$, $\phi \Phi^3$ and
$\phi^4$ which are multiplied by dimensionless couplings. However
we will not include these terms in the lagrangian. The symmetry breaking Lagrangian
${\cal L_{SB}}$ is supposed to contain scale-non-invariant terms,
in particular, a stabilizing (stringy) potential for $\phi$ in the
S-frame. For this reason we write: \beq {\cal L}_{\rm
SB}=-\sqrt{-g} (a \phi^2 + b + c \frac{1}{\phi^2}). \label{SB}
\eeq

Happily, it is possible to satisfy the field equations with
constant values of the fields $\phi$ and $\Phi$ through a proper
choice (but not fine-tuned) values of the parameters
$a, b, c$, maintaining $f<0$ and $\lambda_{\Phi}>0$. We made sure
that $g_s>1$ can be recovered in the equilibrium configuration and
that, consequently, the solution is consistent with the
non-perturbative action that we considered as a starting point.

Here is a possible choice of parameters in string units (see \cite{lettera}):
$f=-2/45$, $\lambda_{\Phi}=0.3$, $a=1$, $b=-\frac{13}{72}$, $c=1/108$, $\zeta=5$. In the equilibrium configuration we have
$\phi_0=\frac{1}{2}$ and $\Phi_0=\frac{1}{3}$.

One more remark is necessary. As already pointed out in \cite{lettera}, the 4-dimensional curvature in the S-frame is {\it constant}. With our choice of parameters we find a positive curvature, $R\simeq10.1$ (in dimensionless units). Therefore we choose the de Sitter metric as our S-frame metric.

\subsection{The scale invariant Lagrangian}
\label{SI}

In this section we will analyze ${\cal L}_{\rm SI}$ following
\cite{Fujii:2002sb, Fujii:2003pa, Zanzi:2010rs, lettera}.

\subsubsection{Classical level - Einstein frame}
\label{vacuum}

The lagrangian
\reflef{bsl1-96}) will be rewritten in the E frame and {\it
spontaneous} breaking of scale invariance will be discovered.

Since our intention is to quantize the theory in the E-frame, we write down the conformal transformation in D dimensions (this will be useful because at this stage we are going to regularize the theory exploiting dimensional regularization). In D=2d dimensional spacetime a scalar field has canonical
dimension (d-1) and the conformal transformation looks like
\begin{equation}
\sqrt{-g}=\Omega^{-D} \sqrt{-g_*}. \label{ctdd}
\end{equation}
In order to recover the usual Einstein-Hilbert term we impose
\begin{equation}
\xi \phi^2=\Omega^{D-2} \label{ctdd}
\end{equation}
that implies
\begin{equation}
\Omega=e^{\frac{\zeta \sigma}{d-1}}, \label{ctdd}
\end{equation}
where the relation $\phi=\xi^{-1/2} e^{\zeta\sigma}$ and $M_p=1$ have been
exploited.

In this way we can rewrite the lagrangian \reflef{bsl1-96}) in the
E frame as
\begin{equation}
{\cal L}_{*}=\sqrt{-g_*}\left( \frac{1}{2} R_* -
    \half g^{\mu\nu}_*\partial_{\mu}\sigma\partial_{\nu}\sigma + {\cal L}_{* matter}
    \right),
\label{eframe}
\end{equation}
where ${\cal L}_{* matter}$ turns out to be
\begin{equation}
{\cal L}_{* matter}= -
    \half g^{\mu\nu}_* D_{\mu}\Phi_* D_{\nu} \Phi_* - e^{2 \frac{d-2}{d-1} \zeta
    \sigma} (\xi^{-1} \frac{f}{4} M_p^2 \Phi_*^2+ \frac{\lambda_{\Phi}}{4!} \Phi_*^4)
\label{lmatter}
\end{equation}
and $D_{\mu}=\partial_{\mu}+ \zeta \partial_{\mu} \sigma$.

The conservation law remains true even after the conformal
transformation, but the symmetry is broken {\em spontaneously} due
to the trick by which a dimensionful constant $M_{\rm P}(=1)$ has
been "re-installed" in \reflef{lmatter}). 

\subsubsection{Anomaly and dilaton coupling}

If we substitute $\Phi_*=v_{\Phi}+\tilde\Phi$ and the expansion
\begin{equation}
e^{2 \frac{d-2}{d-1} \zeta
    \sigma}=1+ 2 \zeta \frac{d-2}{d-1} \sigma +... \label{espansione}
\end{equation}
in formula \reflef{lmatter}), we obtain the following lagrangian
at first order in $\sigma$:
\begin{equation}
-{\cal L}_{1}= 2 \zeta (d-2) \sigma \left( \frac{1}{2} m^2 \tilde
\Phi^2 + \frac{1}{2} \sqrt{\frac{\lambda_{\Phi}}{3}} m \tilde
\Phi^3 + \frac{\lambda_{\Phi}}{4!}  \tilde \Phi^4 \right) -
\frac{3}{4} \zeta (d-2) \sigma \frac{f^2}{\xi^2 \lambda_{\Phi}}
M_p^4 ,  \label{interazioni}
\end{equation}
where $m^2=-\frac{f}{\xi} M_p^2$.
In this way several different 1-loop diagrams can be constructed
(Figure 1): we will consider an explicit calculation starting from
diagram (c).

\begin{minipage}[t]{14.4cm}
%\hspace*{-6.5em}
\baselineskip=0.4em \epsfxsize=14cm
\mbox{}\\[-7.4em]
\hspace*{-2.em} \epsffile{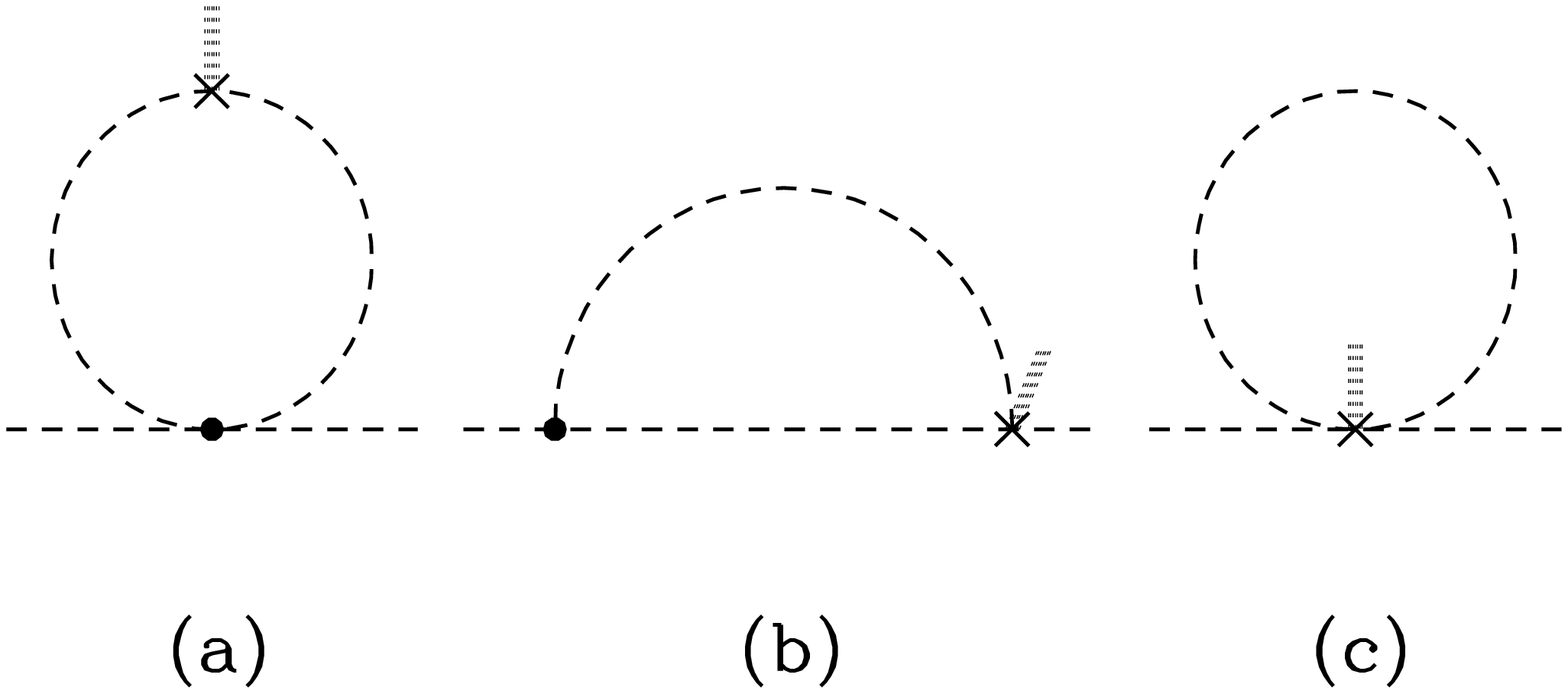}
\mbox{}\\[-5.1em]
Figure 1.  {\sl Examples of 1-loop diagrams for the interaction
$\lmd_\Phi \Phi_*^4$, represented, together with the derived
3-vertex, by a filled circle.
Heavy dotted lines are for $\sigma$. This figure can be found in reference \cite{Fujii:2004bg}.}  \end{minipage}
\label{diagrammi}
\mbox{}\\[-.0 em]

The one-loop diagram is divergent. Exploiting dimensional
regularization the contribution given by the scalar loop is \beq
{\cal I}_{c}=-i \zeta \lambda_{\Phi} (D-4) (2 \pi)^{-D} \int
d^{D}k \frac{1}{\left( k^2 + m^2 \right)}, \label{intloop} \eeq
where the integral in D-dimensions is evaluated explicitly as \beq
I_{c}= \int d^{D}k \frac{1}{\left( k^2 + m^2 \right)} = i \pi^2
(m^2)^{d-1} \Gamma(1-d). \label{intloop} \eeq

Remarkably (2-d) cancels out the pole at d=2 in $\Gamma(1-d)$
leading to a finite result in 4 dimensions:

 \beqa (2-d )\Gamma(1-d )&=&\frac{1}{1-d}(2-d
)\Gamma(2-d )
\nnb\\
&=& \frac{1}{1-d}\Gamma(3-d )\stackrel{d \rightarrow 2}
{\longrightarrow} -1\label{cfrc_4} \eeqa and this is one of the
possible ways in which an anomaly may present itself.

Adding up the contributions from the three diagrams in Figure 1 we
obtain \beqa {\cal I}_{tot}={\cal I}_a + {\cal I}_b+{\cal I}_c=
\frac{1}{\pi^2} \zeta \lambda_{\Phi} m^2\label{vertice} \eeqa or,
in other words, the fundamental (anomalous) interaction vertex
between dilaton and matter \cite{Fujii:2002sb}: \beqa {\cal
L}_{\Phi \Phi \sigma}=-\frac{1}{2} \frac{{\cal I}_{tot}}{M_p}
\tilde \Phi^2 \sigma.\label{vertice} \eeqa

According to the relativistic quantum field theory, the
"anomaly-induced" interaction \reflef{vertice}) with the matter
field $\Phi$ leads us to the contributions depicted in Figure 2.
It seems worthwhile to point out that
the set of diagrams in figure 2 must be considered as an expansion
to {\it all orders} in perturbation theory, namely as a complete
expansion and not as a 1-loop contribution. In other words, since
the interaction vertex between dilaton and matter has been
obtained from a 1-loop calculation (from the conformal anomaly),
every time we add one external leg we are taking into account an
additional loop in the calculation.

\begin{minipage}[t]{14.4cm}
%\hspace*{-6.5em}
\baselineskip=1.7em \epsfxsize=12cm
\mbox{}\\[-0.4em]
\hspace*{-0.1em} \epsffile{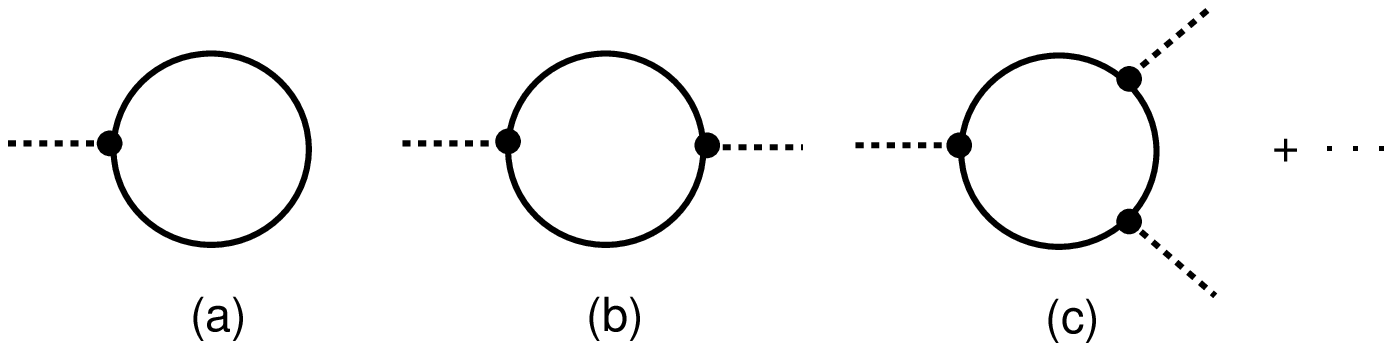}
\mbox{}\\[-0.1em]
Figure 2.  {\sl First three examples of the 1-loop diagrams for
the quantum corrections due to the "anomaly-induced" matter
coupling. Solid and dotted lines are for
$\Phi$ and $\sigma$ respectively. This figure can be found in reference \cite{Fujii:2002sb}.}  \end{minipage}
\mbox{}\\[-.0 em]

\subsection{A stringy solution to the cosmological constant problem}
\label{CC2}

In this section we touch upon the main result of our previous papers \cite{Zanzi:2010rs}: the chameleonic dilaton is parameterizing the amount of scale symmetry of the system in the E-frame. In this way, on cosmological distances scale invariance protects the cosmological constant and the mass of the dilaton.

\subsubsection{Chameleonic scale invariance}
\label{symmetryrestoration}

One crucial element of the analysis discussed in \cite{Zanzi:2010rs} is that scale invariance is almost restored on
cosmological distances in the E-frame and the dilaton
in the E-frame is parametrizing the amount of scale invariance of the problem.

To illustrate this point, let us mention once again that the
quantum dilatonic potential of formula \ref{SB} becomes run-away
after the conformal transformation to the E-frame, because the
conformal factor introduces an exponential suppression. Naturally,
the larger is the value of $\sigma$, the more effective will be
the suppression. Therefore, the question is: {\it what is the VEV
of $\sigma$?} The standard approach to find this VEV would be to
evaluate the effective action of the model exploiting the
deWitt-Vilkovisky method \cite{Vilkovisky:1984st, deWitt:1988dq} (for an introduction on this subject see
for example \cite{Buchbinder:1992rb}). However, in the absence of a detailed analysis,
but inspired by chameleon theories, we
parametrize the length scales of the problem through the value of
$\sigma$. In this way we write $\sigma=\sigma_f(x) + \sigma_b(t)$, where $\sigma_f$ is the fluctuating field and $\sigma_b$ is the background field. We can consider $\sigma_b$ and $\sigma_f$
as two distinct (but related) objects corresponding to different
length scales (see also section \ref{azionieffettive} below). Accordingly, we write $\sigma_b \neq \sigma_f$. Needless to say, $\sigma_b(t)$ is the cosmological (i.e. averaged) field.\\

We are now ready to illustrate the restoration of scale symmetry in the E-frame for large $\sigma$.
\begin{itemize}
\item {\it Constant term in the S-frame}

An explicit
violation of scale invariance on {\it all} distances in the S-frame will be
exponentially reduced in the E-frame, but only for large values of
$\sigma$. In other words, even if our starting point is a S-frame
theory which is {\it not} scale-invariant, a constant term in the
S-frame will be suppressed for large $\sigma$ after the conformal
transformation to the E-frame.

\item {\it The spontaneous breakdown of scale-invariance
introduced by the conformal transformation to the E-frame is
negligible for large $\sigma$.}

As far as the matter sector is concerned, $\Phi$ has finite constant value $\Phi_J$ in the J-frame and the conformal transformation introduces the exponential
rescaling of the $\Phi_*$-field.

As far as the Einstein-Hilbert term is concerned, it is not scale
invariant and we will now show that it is (almost) negligible for
large $\sigma$. This result is based on two elements, namely: (1)
the action \ref{Ltotale} in the S-frame is the result of a quantum (non-perturbative) calculation and (2) quantum
loops can induce a kinetic term for gravitons (i.e. induced
gravity mechanism \cite{Sakharov:1967pk}).
Accordingly, a S-frame kinetic term for gravitons is induced through quantum diagrams. Therefore, the Einstein-Hilbert term in formula \ref{eframe}
is coming from quantum diagrams. However, these diagrams are
suppressed in the E-frame for large $\sigma$,  because {\it all}
interactions are switched off for large $\sigma$
\cite{Witten:1984dg} (i.e. in the weak coupling regime of String
theory). Consequently, the larger is $\sigma$ in the E-frame, the
smaller is the Einstein-Hilbert term.

\item {\it For large $\sigma$, conformal
anomaly is harmless and all the interactions are switched off.} A
scale-dependence of the conformal anomaly has already been
discussed in the literature. For example, in the framework of
AdS/CFT correspondence \cite{Maldacena:1997re} (for a review see
for example \cite{Aharony:1999ti, Maldacena:2003nj}), an
increasing central charge as a function of the energy scale has
already been discussed (the reader is referred to
\cite{Nojiri:1999eg} and references therein). To the best of our
knowledge, this should be a rather general property of conformal
anomalies in agreement with c-theorems (for a review see for
example \cite{Ogushi:2001md}). In our model, this result is
obtained through the exponential rescaling mentioned above. In
more detail, we can write the anomalous coupling \ref{vertice} as:
\beq \frac{1}{2}{\cal I}_{tot} \tilde{\Phi}^2 \sigma =
\frac{1}{2}{\cal I}_{tot} e^{-2 \zeta \sigma} \Phi_f^2 \sigma,
\label{anomaliaexp} \eeq where $\Phi_f$ is the fluctuating
component of $\Phi$ in the S-frame and it is constant. We infer
that for large values of $\sigma$ the coupling is suppressed.

A harmless conformal anomaly for large $\sigma$ will be a crucial element in our analysis, where, as we will show below, large values of $\sigma$ will correspond to large distance scales. As already mentioned in \cite{Zanzi:2010rs}, conformal anomaly is not always harmless in Cosmology. 
In our stringy model, since the dilaton controls the strength of the interactions, {\it all} the interactions ({\it including gravity}) will be switched off for large $\sigma$ (i.e. weak coupling regime). Consequently, we are considering a {\it free} string theory when $\sigma$ is large. For this reason,
in this model we can {\it always} neglect conformal anomaly for large $\sigma$.

\end{itemize}

We infer that for large $\sigma$ scale invariance is restored in the E-frame.
Remarkably, there are a number of consequences of this
restoration of scale-invariance:\\

{\it Consequence 1}.  $\sigma_b$ and $\sigma_f$ acquire different
mass because the degree of divergence of their loop diagrams is
different.\\

{\it Consequence 2}. We extend consequence 1 to all the diagrams
(present of course in infinite number) shown in figure 2.\\

{\it Consequence 3}. For large $\sigma$ the (almost) symmetric
configuration will not be compatible neither with large couplings between dilaton and matter, nor
with large mass scales, including the cosmological constant and the mass of the dilaton. Therefore, the total renormalized vacuum energy
is run-away because scale invariance is restored for large $\sigma$.\\

From these three consequences, since we know that $\sigma$ gets a mass through
the interaction with matter (encoded in formula \ref{vertice}), we infer
that $\sigma_b>\sigma_f$ and, therefore, scale invariance is restored on cosmological distances in the E-frame and $\sigma_b(t)$ is ultralight.
In more detail, the dual mass of the dilaton is obtained through a competition between the run-away branch of the potential and the matter branch. Therefore, the total unrenormalized vacuum energy must be {\it positive}. Had we considered a negative (total) unrenormalized vacuum energy, there would have been no competition between the run-away branch of the potential and the matter branch and this would have clashed with consequences 1 and 2.

Naturally we have an infinite number of different contributions to
the cosmological constant. A vacuum energy term
on short distances does {\it not} contribute to the Dark Energy.
Therefore, the contributions to the vacuum energy which are
relevant for the cosmological constant are {\it global} and not
local. This last comment requires a more detailed discussion.
Expectation values of quantum operators can be rewritten as
spatial integrals weighted by the integration volume (see for
example \cite{Gasperini:2009wp}). On the one hand, if we consider
an integration volume much smaller than the volume of the
Universe, we are dealing with a {\it local} quantity and the
result will be the expectation value of the field (EV).  On the
other hand, if we consider the spatial integral of the field
weighted by the volume of the visible Universe, we are
dealing with a {\it global} (cosmological) object and the result
is the {\it vacuum} expectation value of the field (VEV), which is
the relevant one in the evaluation of the Feynman diagrams
contributing to the cosmological constant. In this way, our
solution to the cosmological constant problem is intrinsically
linked to a {\it dual nature} of the concept of particle. The splitting of a particle in a background (global) part
and in a fluctuating (local) part, that we already discussed for
the dilaton, is extended in our proposal of reference \cite{Zanzi:2010rs} to {\it all the fields} (in particular the E-frame metric has been split in $g^{\mu\nu}_{FRW}+g^{\mu\nu}_{f}$, where $g^{\mu\nu}_{FRW}$ is the Friedmann-Robertson-Walker metric).
Whatever will be the Feynman diagram we consider, if our intention is to evaluate its contribution to the cosmological constant (i.e. to the global renormalized vacuum energy), we must \cite{Zanzi:2010rs} (1) construct the diagram exploiting only the background part of the particles and (2) give a VEV to the external legs (if they are present).

\subsubsection{The chameleonic effective actions}
\label{azionieffettive}

In our proposal of reference \cite{Zanzi:2010rs}, the chameleonic
behaviour of the dilaton is compatible, {\it and it is the
effective result}, of a renormalization process which is carried
on to {\it all} orders in perturbation theory. The success of this
renormalization program is guaranteed by the restoration of
scale-invariance.

It must be stressed that in our approach we do {\it not} evaluate
the diagrams: we exploit the restoration of scale invariance on
cosmological distances to protect the mass of the dilaton and the
cosmological constant. Our argument is valid {\it at all orders}
in perturbation theory.

Remarkably, the chameleonic behaviour of the dilaton and the
splitting of the particles in global and local objects led us to a
formulation of the theory which strongly depends on the choice of
the length scale (parametrized by the value of the dilaton). In particular, as we argued above, large values of
$\sigma$ correspond to large distance scales (i.e. small mass
scales). In this way we can construct two different lagrangians
$L_A$ and $L_B$. On the one hand, the {\it local} particles can be
considered as the degrees of freedom of a lagrangian $L_A$,
describing physical phenomena below an energy scale $M_A$ fixed by
$\sigma_f$. On the other hand, {\it global} particles can be
considered as the degrees of freedom of a lagrangian $L_B$,
describing physical phenomena below an energy scale fixed by
$\sigma_b$ that we call $M_B \simeq H_0 <M_A$. $L_B$ can be
considered as the {\it effective theory} valid in the deep
infrared (IR) region, obtained from the (more fundamental)
lagrangian $L_A$ by integrating out the heavy degrees of
freedom. We infer
that the running of the dilaton towards large values can be
interpreted as the running of the theory towards the IR region
and, in general, a shift of the dilaton towards larger (smaller)
values will correspond to integrating out (in) degrees of freedom
through the exponential prefactor $e^{-\zeta\sigma}$ already
mentioned above. It is time to summarize our chameleonic
lagrangians (in the E-frame).

The local one is written exploiting {\it only local particles} as \cite{Zanzi:2010rs}:
\begin{equation}
 L_{A}=\sqrt{-g_f^*}\left( \frac{1}{2} (R_*)_f -
    \half (g^{\mu\nu}_*)_f\partial_{\mu}\sigma_f\partial_{\nu}\sigma_f + {\cal L}_{* matter} + V_{eff}(\sigma_f)+ ...
    \right)
\label{eframenuova}
\end{equation}
Here the dots include the gauge part of the theory and they might
also include higher derivative (local) gravitational terms.
$V_{eff}(\sigma)$ is the total chameleonic effective potential for
$\sigma$ obtained by adding together (1) a run-away exponential
branch and (2) a matter branch linear in $\sigma$, given by
formula \ref{vertice}. ${\cal L}_{* matter}$ is given by:
\begin{equation}
{\cal L}_{* matter}= -
    \half (g^{\mu\nu}_*)_f D_{\mu}\Phi^*_f D_{\nu} \Phi_f^* - \xi^{-1} \frac{f}{4} M_p^2 (\Phi_f^*)^2+ \frac{\lambda_{\Phi}}{4!} (\Phi_f^*)^4
\label{lmatternuova}
\end{equation}
and $D_{\mu}=\partial_{\mu}+ \zeta \partial_{\mu} \sigma_f$.
Needless to say, $L_A$ is not scale invariant.

The global lagrangian is written exploiting only the background
part of the fields and it is (almost) scale invariant. It is
written as \cite{Zanzi:2010rs}:

\begin{equation}
 L_{B}=\sqrt{-g_{FRW}^*}\left( -\half (g^{\mu\nu}_*)_{FRW}\partial_{\mu}\sigma_b\partial_{\nu}\sigma_b + V_{eff}(\sigma_b)+ ...
    \right),
\label{eframenuova2}
\end{equation}
where the dots represent kinetic terms for massless global gauge
fields. Therefore, the effective lagrangian of the model on
cosmological distances corresponds (basically) to a free
cosmological dilaton $\sigma_b$ in a FRW-background.

\subsubsection{The meV scale}
\label{thecorrect}

In the minimum of the effective potential we can write \beq
V(<\sigma_b>)= \rho_m B(<\sigma_b>), \eeq where $\rho_m$ is the
matter energy density and $B(\sigma)$ is the usual function of a
chameleonic model where the coupling is encoded. In the case of the coupling
\ref{vertice}, we identify the matter energy density in the
E-frame with $\frac{1}{2}m^2 \tilde{\Phi}^2$. Therefore we have
\beq B= \frac{1}{M_p \pi^2} \zeta \lambda \sigma \eeq and if we
choose $\lambda \simeq \zeta \simeq 1$, the condition $B=1$ (necessary to recover the correct DE scale through the chameleon mechanism)
requires $\sigma \simeq 10$. In the absence of a detailed analysis,
we have no theoretical grounds to support this value of $\sigma$.
Happily, however, this choice of parameters and fields is not a
fine-tuning and the B-function is linear, therefore, we can tolerate a certain
deviation of the value of $\sigma$ in the minimum from the
planckian scale.

Remarkably the meV-scale is recovered and the cosmological constant is under control {\it only} in the E-frame (which is selected to be the physical frame). The classical equivalence between String and Einstein frames is lost at the quantum level (for a more detailed discussion, the reader is referred to \cite{Zanzi:2010rs, lettera}).

\setcounter{equation}{0}
\section{Intermediate conclusions (some open problems)}
\label{openp}
In the previous section we touched upon the construction of the model previously discussed in \cite{Zanzi:2010rs, lettera}. Summarizing, after dilaton stabilization is achieved in the S-frame in the strong coupling region of string theory, a conformal transformation to the E-frame modifies drastically the dynamic of the system: a chameleonic behaviour of the dilaton is obtained exploiting a competition between a run-away potential (which reflects the restoration of scale symmetry on cosmological distances) and a matter branch (which is the result of a conformal anomaly). The Einstein-Hilbert term, which is {\it not} scale-invariant, is suppressed at large $\sigma$ because we generate the kinetic term for gravitons through the induced gravity mechanism. This means, in particular, that the space-time itself is chameleonic. In other words, in our proposal, spacetime has a hive-structure: it is locally dynamical, but globally (almost) non-dynamical and these two facts do not clash with each other.\\  Another crucial element of the analysis is the dual nature of the concept of particle: all the fields are split in a global component and in a local one. This splitting is relevant in the construction of the effective action. It seems worthwhile to add a few comments about this point we mentioned last. Substituting the splitting $\sigma=\sigma_b(t)+\sigma_f(x)$ in the dilatonic kinetic term we can write $\partial \sigma \partial \sigma= \partial \sigma_b \partial \sigma_b + 2\partial \sigma_b \partial \sigma_f +\partial \sigma_f \partial \sigma_f$ where the first term is the subdominant one because, as already mentioned in the previous section, $\sigma_b(t)$ is ultralight. This means that locally (e.g. in this room), even if the dilaton is the sum of a local fluctuating component and of a background (global) one, only the fluctuating component is relevant, because the global dilaton has a negligible kinetic term and it is non-physical. This set-up opens up some problems, for example:
\begin{itemize}
\item the careful reader may be worried about a chameleonic behaviour of the string dilaton, because a shift of $\sigma$ could induce a shift in the fundamental constants. For example, the density gap between the atmosphere and a common solid object could induce an unacceptable shift of the string coupling $g_s$ and therefore of the fundamental constants.
\item The dual nature of the particles looks unusual and should be further discussed.
\item As far as the quantization of the model is concerned, we proceeded stepwise discussing three different quantization steps. {\it Step 1}, in the S-frame the quantum contributions are encoded in the ansatz for the form factors. {\it Step 2}, another quantization is given by the diagrams of Fig. 1, where dimensional regularization is exploited in the E-frame and conformal anomaly is discovered. {\it Step 3}, one last quantization is evaluated starting from the anomaly induced interaction vertex (see Fig. 2). The restoration of scale invariance on cosmological distances in the E-frame led us to a very peculiar QFT: the result of a loop calculation is not constant, but it is a function of the length scale. From the elements we gathered we do not manage to infer the details of the regularization mechanism for the third quantization step. Indeed, in principle, many regularization scenarios are allowed, because at this stage the only condition that they must satisfy is to be compatible with the chameleonic scale invariance. This point should be further discussed. As already mentioned in the introduction, one first possibility is to introduce in the model a simple constant cut-off and to imagine that, for some reason, the larger is the matter density, the larger is the degree of divergence of the loops. In particular we can imagine a quadratic divergence for a loop constructed with local particles, but a logarithmic divergence for the same loop constructed with global particles. A different possibility is to imagine that the functional dependence of the loops on the cut-off is fixed, but, for some reason, the cut-off is a function of the length scale. It would be rewarding to solve this ambiguity between the two scenarios and to make the regularization mechanism clear and easy to understand. Consequently, the question is: {\it what kind of regularization mechanism do we use in our step 3?}
\end{itemize}

These problems will be solved (at least partially) in the next sections. 

\setcounter{equation}{0}
\section{A partial solution to the problems}
\label{resolution}

We will now start in section \ref{GV} describing a model recently proposed by G. Veneziano \cite{Veneziano:2001ah} to saturate the fundamental couplings in the strong coupling region of string theory. We will proceed further in section \ref{CS} by creating a link between the Veneziano's model and our model of section \ref{CC} and, as a final step, we will discuss once again the three problems mentioned above.

\subsection{The Veneziano's model}
\label{GV}

\subsubsection{The claim}
\label{theclaim}
Consider a  toy model of gauge and gravitational
interactions in $D \ge 4$ space-time dimensions,
 minimally coupled to a large number 
of  spin $0$ and spin  $1/2$ matter fields. 
Let us endow the model with a cut off $\Lambda$,
assumed to be finite and to
preserve gauge invariance and general covariance. 
Let us also neglect, for the moment,  matter self interactions.
The tree-level action of the model is
\bea
S_{0} &=& - {1 \over 2} \int d^{D}x \sqrt{-g_*} 
\left[ \kappa_{0}^{-2} R_* + {1 \over 2} g_{0}^{-2} 
\sum_{k=1}^{N_1}  
F^{k}_{\mu\nu} F^{k\mu\nu} \right] 
+  \sum_{i=1}^{N_{0}}  
\left(  (D_{\mu} \Phi_*)_{i} (D^{\mu} \Phi_*)_{i}
 + m_0^2 \Phi_{*i}^2 \right) \nonumber \\
      &+& \int d^{D}x \sqrt{-g_*} \left[ \sum_{j=1}^{N_{1/2}} \left( \bar{\psi_*}_{j} 
(i\gamma \cdot D \psi_*)_{j}
 + m_{1/2} \bar{\psi_*}_{j} \psi_{*j} \right) 
 + ~ \dots 
\right] ,
\label{treeaction}
\eea
where  dots stand for an ultraviolet completion of the model
 implementing the UV cutoff. We have given for simplicity a 
common mass $m_0$ to all
the spin zero fields and a common mass $m_{1/2}$ to all spin $1/2$ fields.
These masses are assumed to be small compared to the UV cut-off $\Lambda$.

The claim of the paper \cite{Veneziano:2001ah} is related to the renormalized value of the 
gauge and gravitational couplings, $g^2$ and 
$\kappa^{2} = 8 \pi G_{N}$, as a function of their
bare values, $g_0^2$ and $\kappa_0^{2}$, and of $\Lambda$, when the total number 
of matter fields $N = N_{0} + N_{1/2} \rightarrow \infty$, 
while  their  relative ratios are kept fixed.
In particular, the claim can be summarized as follows: in the above-defined model (supposed to be a UV finite higher dimensional theory of all interactions provided by superstring theory) modulo
a certain generic assumption  about one-loop contributions, the 
following bounds hold:
\bea
\alpha_g \equiv g^2 \Lambda^{D-4} &<& {c_1 \over N^p} \; , D > 4\;,
 \nonumber \\
\alpha_G \equiv G_N \Lambda^{D-2} &<&  {c_{2} \over N} \; ,  D \ge 4\;,
\label{bound}
\eea
where $c_{1}, c_{2}$ are positive constants (typically  smooth functions
of the relative abundances $N_{i}/N$) to be computed 
at the one-loop level, and $p$ is a number between
$0$ and $1$. The case of the gauge coupling at $D=4$
 needs a separate discussion because of infrared effects.

Moreover, both bounds are saturated in the compositeness 
(infinite-bare-coupling) limit and, therefore, this limit can be realistic.

\subsubsection{The proof}
\label{theproof}
Consider the Feynman  path integral corresponding to the action 
(\ref{treeaction}) and integrate out completely (i.e. on all
scales) the matter fields. In this way we take into account the renormalization due to matter fields. We remove the E-frame star index ($*$) from the formulas of the proof in order to simplify the notation: in this paragraph $g_{\mu\nu}$ is the E-frame metric (we will come back to our standard notations in the last paragraph).  The result in the E-frame is:
\bea
I &=& \int  dg_{\mu\nu} d A_{\mu}^k d \Phi_{i} d \psi_j
 {\rm exp} \left(i (S_{0, {\rm gravity}} +  S_{0, {\rm gauge}} +
 S_{0, \rm matter} + \dots) \right) \nonumber \\
  &=& 
\int  dg_{\mu\nu}  d A_{\mu}^k {\rm exp} (i S_{{\rm eff}}) \; , 
\label{Fintegral}
\eea
where
\beq
 S_{{\rm eff}} = S_{0, {\rm gravity}} + S_{0, {\rm gauge}}
 - 1/2  ~~ {\rm tr~log \nabla^2} (g, A) +  
  {\rm tr~log} (\gamma \cdot D (g, A))  
 \; , 
\label{Seff}
\eeq
and the trace includes the sum over the representations to which
the matter fields belong, in particular a sum over ``flavour" indices.

The result contains local as well 
as non-local terms. In  $D>4$ the non-local terms start with at least 
four derivatives. In $D=4$ this is still true for the gravity
part but non-local contributions appear already in the
$F^2$ terms of the gauge-field action. Thus we write:
\beq
  S_{{\rm eff}} =  S_{0, {\rm gravity}}\left(1 + 
(c_0 N_0 + c_{1/2} N_{1/2})  \kappa_0^2 \Lambda^{D-2} \right)
+ \nn \\ + S_{0, {\rm gauge}} \left(1 + g_0^2 (\beta_0 + \beta_{1/2}) {\Lambda^{D-4} \over
D-4} 
\right) +  S' ,
\label{Seffexp}
\eeq
with the following explanatory remarks.
 $S'$ contains higher derivative (and generally non local) terms.
 The other terms in the  gauge plus gravity effective action are
local with the already mentioned exception of the gauge kinetic term
in $D=4$: this is indicated symbolically by the presence of a 
pole at $D=4$, to be explained better below.
The constants 
$c_0, c_{1/2}, \beta_0, \beta_{1/2}$ are in principle computable
 in any given theory; their order
of magnitude in the large $N$ limit will be discussed below.

We now have to discuss the effect of including gauge and gravity loops or,
if we prefer, to complete the functional integral by integrating over
$g_{\mu\nu}$ and $A_{\mu}^k$ after having introduced suitable
sources. Appropriate large-$N$ limits can help. Indeed, if we consider a large-$N$ limit such
that, not only $N_0, N_{1/2} \rightarrow \infty$, but also
$\beta_0  + \beta_{1/2} \rightarrow \infty$, the effective coupling after
matter-loop renormalization is arbitrarily small and the 
one-gauge-loop contribution dominates the remaining functional integrals (the theory
having become almost classical).  Let us start with
the effect of these large-$N$ integrations on the gauge kinetic term, i.e. with
the renormalization of the gauge coupling due to gravity and gauge loops.
Obviously, such a renormalization adds to the one due to matter loops, and
already included in (\ref{Seffexp}). 
The final low-energy gauge effective action is \cite{Veneziano:2001ah}
\beq
\Gamma_{eff}^{gauge} = - {1 \over 4} \int \sqrt{-g} 
\left[  g_{0}^{-2} + (\beta_0  + \beta_{1/2})
 {\left(\Lambda^{2 \epsilon} -
(q^2 + m^2)^{\epsilon} \right)\over \epsilon} - 
\beta_1 {\left(\Lambda^{2 \epsilon} -
(q^2)^{\epsilon} \right)\over \epsilon}  \right]  F_{\mu\nu}^2 \;,
\label{Gammagauge}
\eeq
 where $\epsilon = (D-4)/2$ and, for the sake of notational simplicity,
 we have taken $m_0 = m_{1/2} = m$.

In order for the approximations to be justified we need to argue that
the quantity $(\beta_0  + \beta_{1/2})$ is sufficiently large and positive, 
indeed that it is parametrically 
larger than the gauge field contribution $\beta_1$. The latter
is proportional to the quadratic Casimir of the adjoint representation $C_A$.
This condition is satisfied
provided $\beta_0  + \beta_{1/2} \sim C_M N_f \frac{d_M}{d_A} \gg C_A$, 
where $d_M, C_M$ represent  dimensionality
and  quadratic Casimir for the matter representation $M$, respectively.  In a QCD-like theory
with gauge group $SU(N_c)$ this would correspond to a large $N_f/N_c$ ratio.

\begin{itemize}
\item {\it Renormalization of the gauge coupling.} What happens to the effective theory at low energy depends on $D$.
We here consider only the $D=4$ case and we refer the reader to the original paper \cite{Veneziano:2001ah} for a more complete analysis. The poles in 
(\ref{Gammagauge})
at $D=4$  have to  be interpreted as infrared logarithms containing in their argument
the box operator (as well as the mass for the matter contribution). They
  provide the well-known logarithmic running of the gauge couplings.
In this case the  bound on the gauge coupling is scale dependent and reads \cite{Veneziano:2001ah}:
\bea
4 \pi \alpha_g(q) &\sim& [g_{0}^{-2}  + 
(\beta_0  + \beta_{1/2}) \log(\Lambda^2/(q^2 + m^2)) - 
\beta_1 \log(\Lambda^2/(q^2)]^{-1}   \nonumber \\
&\le& 
\left[(\beta_0  + \beta_{1/2}) \log(\Lambda^2/(q^2 + m^2)) - 
\beta_1 \log(\Lambda^2/q^2)\right]^{-1} ~~, ~~ D = 4 \;,
\label{alphab4}
\eea
 and, again generically, the limit is reached at infinite bare coupling. 

\item{\it Renormalization of the Newton constant.} Here the situation
is the same at all $D \ge 4$. The low-energy action is local and, since 
 gravity couples equally to all fields, the constants $c_0, c_{1/2}$ are just
some calculable $N$-independent  numbers. Since the quantity
$(c_0 N_0 + c_{1/2} N_{1/2})$ is assumed to be parametrically large, graviton-loop
contributions are subleading at large $N$. The gauge field contribution can be estimated accurately
in the one-loop approximation. The final result is \cite{Veneziano:2001ah}:
\beq
\Gamma_{eff}^{gravity} = - {1 \over 2} \int \sqrt{-g} 
\left[  \kappa_{0}^{-2} + (c_0 N_0 + c_{1/2} N_{1/2} + c_1 N_1) \Lambda^{D-2}
  \right] R \;,
\label{Gammagrav}
\eeq
where $N_1$ is the total number of gauge bosons (the dimensionality of the
adjoint representation) and $c_1$ is also a calculable number of $O(1)$.

We thus obtain the bound on the effective Newton constant:
\beq
8 \pi \alpha_G \equiv \kappa^{2} \Lambda^{D-2} \sim
 [\kappa_{0}^{-2} \Lambda^{2-D} + 
(c_0 N_0 + c_{1/2} N_{1/2} + c_1 N_1)]^{-1} \le
 (c_0 N_0 + c_{1/2} N_{1/2} + c_1 N_1)^{-1} ,
\label{GNbound}
\eeq
where, once more, the bound is saturated in the compositeness limit, $\kappa_{0}
 \rightarrow \infty$\end{itemize}

In the next section we will create a link between the Veneziano's model and the chameleonic one discussed in section \ref{CC}. 

\subsection{Cosmic strings, fundamental constants and stringy regularization}
\label{CS}

Let us come back to our model of section \ref{CC}. We start by {\it assuming} that the form factors in the strong coupling region of string theory are able to guarantee in the E-frame a ''tree-level'' dependence of the couplings on the dilaton $\sigma$: \bea (\frac{M_p}{M_S})^2=e^{2\zeta\sigma}=\frac{1}{g_0^2}=\frac{1}{g_s^2}.\label{albero}\eea 

Let us introduce in the model a large number of species.
We now connect the tree-level action of the Veneziano's model to the first lagrangian that we wrote in the E-frame (and remarkably that is a {\it quantum} lagrangian because it is obtained from the S-frame lagrangian which encodes quantum loops in the form factors). In order to establish the connection we write:

\bea
L_{E-frame} &=& \sqrt{-g_*} 
\left[ \kappa_{0}^{-2} R_* - \half g^{\mu\nu}_*\partial_{\mu}\sigma\partial_{\nu}\sigma  -
    \half g^{\mu\nu}_* D_{\mu}\Phi_* D_{\nu} \Phi_* - m_0^2 \Phi_*^2 - \lambda \Phi_*^4 \right] \nonumber \\ 
&+& \sqrt{-g_*} \left[  \frac{1}{g_0^2(\phi)} F^2 -[\xi^{-2} (a  e^{-2\zeta\sigma} \xi +  b \xi^2 e^{-4\zeta\sigma}+ c \xi^3 e^{-6\zeta\sigma})] 
\right]\nn\\&\equiv& L_{Veneziano} ,
\label{treeactionbis}
\eea
and $D_{\mu}=\partial_{\mu}+ \zeta \partial_{\mu} \sigma$.

At this stage, as far as the quantization procedure is concerned, only step 1 has been accomplished. The non-minimal coupling term has been obtained through quantum diagrams and we wrote the conformal transformation
\bea \xi \phi^2=M_p^2 e^{2 \zeta \sigma}.
\label{trconf}
\eea

Let us now come back to the couplings of the model.
We wrote the ratio $\frac{M_p}{M_S}$ and $g_0$ (the gauge couplings) with a string ''tree-level'' dependence on the dilaton: \bea (\frac{M_p}{M_S})^2=e^{2\zeta\sigma}=\frac{1}{g_0^2}=\frac{1}{g_s^2},\label{albero2}\eea where this formula is a direct consequence of the assumed structure of the form factors\footnote{Considering the structure of the lagrangians and the stringy nature of the model, the ansatz \ref{albero} about the couplings seems natural. We will further analyze the couplings of the model in a future work.}. Remarkably, at this stage of the quantization procedure, the couplings are a decreasing function of $\sigma$ and the form factors guarantee that in the E-frame formula \ref{albero} is valid not only in the weak coupling regime, but also in the strong coupling one.
From \ref{albero} and \ref{trconf} we infer that the $\phi$ field has dimensions given by \bea [\phi]=\frac{M_p^2}{M_s}. \label{dimensioni}\eea Interestingly, the dimensionless field $\phi/M_s$ is parametrizing the ratio $(M_p/M_s)^2$. Therefore, our S-frame dilaton, which is fixed to be smaller than one, is compatible with strings smaller than the Planck length. This comment might open up connections with transplanckian scattering (in a constant curvature space), however, we leave for the future a careful analysis of these ideas that will start from \cite{Ferrari:1988cc, Podolsky:2000cx, Podolsky:2002nn, Podolsky:libro1, Podolsky:libro2} and related references.

To proceed further, we quantize once again the theory (see Fig. 1), this is the so-called {\it step 2} (see our intermediate conclusions but also reference \cite{Zanzi:2010rs}) and we exploit the results of section \ref{GV} to infer that:
\begin{itemize}
\item {\it If the Veneziano's mechanism can be exploited locally in this model}, we can write A) $\alpha_G \simeq (M_S/M_p)^2 \simeq \frac{1}{N} \simeq constant$ (see formula \ref{GNbound}) granted that the ratio $(M_p/M_S)^2$ after step 1 is subdominant with respect to the contribution involving the number of species and also B) $\alpha_g(\sigma)\simeq constant$ (see formula \ref{alphab4}). Formula (A) might be useful to keep under control the local variations of the gravitational coupling and, moreover, it is compatible with a chameleonic behaviour of the string length: a shift in the value of the chameleonic dilaton implies a shift in the Planck mass (which gets a monotonic contribution from step 1, see also \cite{lettera}), namely, see formula (A), a shift in the string length. If we now consider formula B, we see that it might be useful to keep locally under control the dangerous variations of fundamental gauge couplings. As already mentioned, more research efforts are required to clarify whether the Veneziano's mechanism can be exploited locally in this model.
\item{\it Globally.} In the IR-region \ref{albero} is certainly valid. One remark is in order. There is no clash between a loop-induced Planck mass and formula \ref{albero}. As already mentioned above, the S-frame dilaton is fixed in the equilibrium configuration (which is located in the non-perturbative regime), therefore, the non-minimal coupling term $\phi^2 R$ is the result of a {\it quantum} calculation and its presence in the S-frame lagrangian {\it does not depend on the E-frame dilaton}. With this explanatory remark, in this paper we call \ref{albero} the string {\it ''tree-level''} formula, keeping in our mind that 1) the non-minimal coupling term is a quantum result, 2) the Planck mass is loop-induced and 3) \ref{albero} is certainly valid in the weak coupling regime of the E-frame. These three elements do not clash with each other. Therefore we have that 1) interactions are switched off globally; 2) in the IR region, $\sigma$ has a large value (hence $M_p>>M_S$), but we know that $M_p$ is small because it is obtained through the induced gravity mechanism (see also \cite{lettera}). Consequently, $M_S$ is very small in the infrared. Formula \ref{albero} is compatible with a chameleonic behaviour of the string length.
\end{itemize}

Summarizing, the string length is chameleonic: the string mass is an increasing function of the matter density and, consequently, global particles are cosmic strings (this issue will be further analyzed below).
The chameleonic behaviour of the string length is the result of various elements:\\ a) non-increasing couplings as functions of $\sigma$ (the formula $M_p/M_S=constant$ and the ''tree-level'' relation \ref{albero} are both compatible with a chameleonic string length);\\ b) the induced gravity mechanism with a decreasing Planck mass as a function of $\sigma$ (see \cite{Zanzi:2010rs, lettera});\\ c) the chameleonic behaviour of the dilaton which parametrizes the amount of scale symmetry of the system in the E-frame.\\ 
It seems worthwhile to point out that the mass of global matter particles and their potential chameleonic behaviour depend on the function $M_p=M_p(\sigma)$, namely, on the induced gravity mechanism. Remarkably, the matter fields are removed from the global lagrangian through the exponential factor even if they are not heavy. In section \ref{debroglie} we will point out the chameleonic behaviour of matter fields and we will further discuss these issues. A careful phenomenological analysis is definitely necessary.

A chameleonic string length is not only fully consistent with the chameleonic scale invariance of the model, but it also sheds some light on the regularization mechanism of the third quantization step of the theory. The cut-off is a function of the chameleonic dilaton: this is a stringy regularization mechanism where the cut-off is the (dilaton-dependent) string mass. When $\sigma=\sigma_f$ (i.e. locally), the model is characterized by small interacting strings and by Feynman diagrams with a large cut-off. On the contrary, when $\sigma=\sigma_b$ (i.e globally), strings are cosmic, almost non-interacting and the UV cut-off is smaller than before. We can roughly estimate this quantity we mentioned last. Indeed, globally we can write the critical density $\rho_c$ as (see formulas \ref{albero},\ref{dimensioni})
\bea
\rho_c \simeq \rho_m \simeq m_*^2 \Phi_*^2 \simeq (M_p^2 \phi^2 e^{-2 \zeta \sigma})_{global} \simeq (M_p^2 \frac{M_p^4}{M_s^2} e^{-2 \zeta \sigma})_{global} = (M_p^{global})^4,
\label{critic}
\eea
where we exploited the fact that in the S-frame we have $\phi \simeq \Phi$. Moreover, globally we can write (see formulas \ref{albero}, \ref{critic})
\bea M_s^{global} \simeq \frac{M_p^{global}}{e^{\zeta \sigma_b}} \simeq \Lambda_{DE} e^{-\zeta \sigma_b}, \eea
where $\Lambda_{DE}$ is the meV Dark Energy scale.
Therefore, global particles are expected to be cosmic strings in this model.

The importance of a finite string size in connection to UV regularization has been already discussed by Kempf and Mangano in the framework of non-commutative geometry (see \cite{Kempf:1996nk}). Therefore, in our paper a remarkable connection between chameleon theories, (cosmic) strings and non-commutative geometry is pointed out. More research efforts are necessary to explore the potential consequences of these connections.

Interestingly, in this model a large hierarchy between the Planck scale and the string one can be present. This is true, on the one hand, globally (because the string coupling vanishes), on the other hand, it remains true if the Veneziano's mechanism can be exploited locally in this model (because of the large number of species - see formula \ref{GNbound}). The idea of a small string scale is not new and potential connections between our model and the existing literature on the subject should be carefully investigated, in particular:
\begin{itemize}
\item potential connections with Large Volume Compactifications \cite{Balasubramanian:2005zx}.
\item potential connections with higher spins theories \cite{Fang:1978wz, Fronsdal:1978rb, Sagnotti:2003qa}.
\item potential phenomenological signatures of the model, see for example \cite{Antoniadis:2011bi}.
\end{itemize}

Another interesting research project should be mentioned, namely, the potential connection between the shortest length scale of nature, the number of species and long distance symmetries (i.e. the amount of scale invariance in the IR region) of our model. In order to avoid confusion, we point out once again that the relevant length scale in this research direction is the shortest possible length scale of nature, it is {\it not} the global UV cut-off. The question that we will try to address in the future is whether it is also possible to link $\rho_c$ (i.e. the cosmological constant) to the shortest possible length scale of nature. Some clues of this remarkable link are present in our model and we discuss them in section \ref{mixing}, however, we leave for the future a careful theoretical investigation that will start from \cite{Dvali:2007hz, Dvali:2010vm, Susskind:2005js, Ghirardi:1997xx, Giddings:2007pj, Brustein:2009ex} and related references.

\setcounter{equation}{0}
\section{Chameleonic matter fields}
\label{debroglie}

In the framework of the model of section 2, let us consider the E-frame lagrangian and let us discuss separately the kinetic term and the mass term of matter particles.
\begin{itemize}
\item As far as the mass term is concerned, we notice that the mass of matter particles is related to the Planck mass and it is chameleonic. 
\item The kinetic term of matter fields is chameleonic. Indeed, the terms of $D_{\mu} \Phi_*$ are combinations of $\partial \sigma$, $\partial \Phi_*$ and $\Phi_*$ which are all chameleonic. 
\end{itemize}
To proceed further, we point out that, on the one hand, the global dilaton is homogeneous and non-stationary, while, on the other hand, the local dilaton is non-homogeneous and stationary. Since we write the matter fields in the E-frame as $\Phi_* \propto e^{-\zeta \sigma}$, we infer that these dilatonic properties are induced on the matter fields. Consequently, locally (i.e. in this room), the time-derivative of the matter fields is subdominant with respect to their spatial gradient. Hence, the mass-squared of a local matter particle is basically its 3-momentum, because we can write
\bea
g^{\mu\nu}_f p_{\mu} p_{\nu} \simeq g^{ij}_f p_i p_j, 
\eea
where $i$ and $j$ are spatial indexes while $\mu$ and $\nu$ are space-time indexes.
We infer that, in this model, the de Broglie wavelength of local matter particles $\lambda= h/p$ is chameleonic.

The careful reader may be worried by these ideas and may ask a number of questions.\\
First of all, if we measure (locally) the mass of the (local) electron, we should obtain 0.5 MeV and this result should not depend on the matter density of the environment, namely, the mass of the electron is the same in the air of this room and inside a wall of this room. Therefore, a chameleonic behaviour of standard matter particles looks unacceptable at first sight. However, the result of a measurement is always a dimensionless ratio. For example, if our intention is to measure the mass of the electron, the result of a measurement is the ratio between the mass of the electron and a reference standard mass scale that, in the E-frame, can be reasonably chosen as the Planck mass. We infer that a chameleonic shift of the electron mass seems to be unobservable at this stage, because it is linked to the same shift of the standard reference (Planck) mass.  One interesting line of development will further investigate potential consequences of chameleonic matter fields in this model.  \\
Another interesting question comes from the conformal anomaly: it produces a linear term in $\sigma$ which multiplies the matter density and, therefore, the mass (squared) of the matter fields. The question is whether this linear term can have an observable impact on the measurement of the mass of the matter particles. We leave for the future a careful phenomenological investigation. However, in this paper we point out that even if we include the linear term in the evaluation of the mass of a matter particle, the dependence on $\sigma$ will be similar to the dependence of the Planck mass on $\sigma$. Indeed, large (or at least not small) values of $\sigma$ are necessary if our intention is to create a reasonable matter density starting from a large planckian energy density. Therefore, the exponential factor in the Planck mass is the dominant one. Consequently, \bea \frac{M_p(\sigma)}{m_{\Phi_*}} \simeq constant. \eea

\setcounter{equation}{0}
\section{Towards a UV-IR mixing}
\label{mixing}

In this paragraph we are going to touch upon some ideas that we consider particularly interesting as a future research direction, namely, the potential connection between the shortest length scale of nature, the number of species and long distance symmetries (i.e. the amount of scale invariance in the IR region) of our model. 

One of the elements we exploited in our discussion of the chameleonic behaviour of the string length is the dependence of the gravitational effective coupling $\alpha_G$ on the dilaton $\sigma$. In general, $\alpha_G$ is determined not {\it only} by the Planck mass, but also by the momentum-transfer $p$. In particular we can write $\alpha_G (p^2) \equiv 16 \pi G_N p^2$. In our paper the momentum transfer is bounded by the string mass (which is the UV cut-off). We explicitly discussed two different regimes: 
\begin{itemize}
\item The string ''tree-level'' relation \ref{albero}.
\item The Veneziano's regime. $\alpha_G \simeq (\frac{M_S}{M_p})^2 \simeq 1/N \simeq constant$.
\end{itemize} 
Even if these two regimes are both compatible with a chameleonic string length, they are very different from each other. Remarkably, the result of a measurement of distance is a dimensionless ratio of two mass scales. In other words, if our intention is to measure the distance between two points, two different length scales must be considered: 1) the distance between the two points; 2) the length of a ''ruler'' that we use as a reference length. Therefore, in a particle physics language, we can say that two different mass scales are present and the result of a measurement of distance is a dimensionless ratio between these two mass scales. In our model we probe short distances with strings (i.e. through $M_s$), while the reference length scale in the E-frame is $M_p^{-1}$. Therefore, as far as the shortest length scale of nature is concerned, a chameleonic shift of the string length in the Veneziano's regime can be considered trivial. On the contrary, a shift in the number of particle species is highly non-trivial in connection to the shortest length scale of nature: if we change $N$, we change the ratio $M_S/M_p$ and a different physical configuration is obtained. To proceed further, a junction between the various regimes is required, i.e. a continuity of the function $\alpha_G(\sigma)$. This might be the source of a UV-IR mixing. Consequently, a shift in the number of particle species might be connected to the cosmological constant: if we change $N$, a different junction condition must be respected and, consequently, different parameters should appear in the function $\alpha_G(\sigma)$. The final outcome might be a connection between the shortest length scale of nature (UV) and the cosmological constant (IR). More research efforts are required to make this scenario clear.

\section{Conclusions and possible lines of development}
\label{conclusions}

The chameleonic behaviour of the dilaton and of the spacetime structure has already been discussed in reference \cite{Zanzi:2010rs}. In this article we further illustrated our proposal discussing also some new results. In this paper we proceeded stepwise: \\ 1) we pointed out that the value of the chameleonic dilaton in the Einstein frame parametrizes also the string length, in particular, the string mass is an increasing function of the matter density;\\ 2) global particles are simply cosmic (and almost non-interacting) strings. Consequently, the dual nature of the concept of particle introduced in \cite{Zanzi:2010rs} is clarified and easy to understand.\\ 3) In the last quantization step, in the Feynman diagrams of our model, the UV cut-off (which is chosen to be the string mass) is a function of the chameleonic dilaton. This is a stringy regularization mechanism compatible with the chameleonic scale invariance (and some of its aspects resemble the regularization discussed by Kempf and Mangano in the framework of non-commutative geometry \cite{Kempf:1996nk}). \\4) We showed that a large number of particle species might be useful to keep locally under control the dangerous variations of the fundamental constants, however, more research efforts are necessary to make this point clear.\\5) The chameleonic behaviour of matter fields and of the de Broglie wavelength of local matter particles has been pointed out.\\ 6) We briefly touched upon some ideas (which are still waiting for a detailed analysis and full confirmation) regarding a potential connection between the shortest length scale of nature and the cosmological constant. However, a detailed phenomenological analysis of the entire model is required to test these ideas.\\ Many lines of development are currently viable. An incomplete list includes:\\ A) gravity is totally absent globally and a detailed cosmological analysis is definitely necessary to guarantee that this fact does not clash with cosmological constraints.\\ B) The potential consequences of a small ratio $M_S/M_p$ should be discussed in connection to Large Volume Compactifications \cite{Balasubramanian:2005zx} (ii) higher spins theories \cite{Sagnotti:2003qa} and (iii) collider experiments \cite{Antoniadis:2011bi}. \\ C) A different research project is given by the potential connection between chameleon fields, particle species, black-hole physics, shortest possible length scale of nature and long-distance symmetries: this line of development should be further investigated starting from \cite{Dvali:2007hz, Dvali:2010vm, Susskind:2005js, Ghirardi:1997xx, Giddings:2007pj, Brustein:2009ex, Griguolo:2001ez} and references therein. Particular attention should be dedicated to the potential connection between the cosmological constant and the shortest possible length scale of nature.\\ D) The connection between chameleon theories, cosmic strings and non-commutative geometry should be further investigated starting from \cite{Kempf:1996nk, Griguolo:2001wg, Griguolo:2001ez, Vilenkin:libro, Erdmenger:2009ll, Seiberg:1999vs} and related references. \\E) The dependence of the saturation mechanism of the couplings on the details of the particle content of the theory and on the matter self-interactions should be analyzed.\\ F) Potential connections with transplanckian scattering (see formula \ref{dimensioni}) should be discussed starting from \cite{Ferrari:1988cc, Podolsky:2000cx, Podolsky:2002nn, Podolsky:libro1, Podolsky:libro2} and related references.\\ G) The anomaly-induced vertex $\Phi_*^2 \sigma$ (i.e. matter-matter-dilaton) resembles the QED interaction vertex (i.e. matter-matter-photon). Therefore, a chameleonic bremstrahlung should be present. In particular, in this model, an accelerated matter particle looses its energy through emission of dilatons. In a proper experimental situation (e.g. in a collider) this energy loss might be detectable and it would be an indirect evidence of a chameleonic dilaton. This possibility should be investigated. \\
H) In this model we deal in a new way with the wave-like nature of matter particles. Remarkably, the string length and the de Broglie wavelength of local matter particles are {\it both} chameleonic. This comment seems to suggest at this stage a scenario where the probability waves of matter particles are interpreted as waves on quantum matter strings. We think that this idea deserves a careful theoretical investigation. An interesting line of development will try to give firm theoretical grounds to this idea that, at this moment, we are not able to support.\\
I) The chameleonic behaviour of matter particles should be further analyzed and particular attention should be dedicated to its potential phenomenological signatures.\\
L) The potential identification of the dilaton with the Higgs field should be discussed starting from \cite{Foot:2008tz} and related references.\\

%%%%%%%%%%%%%%%%%%%%%%%%%%%%%%%%%%%%%%%%%%%%%%%%%%%%%%%%%%%%%%%%%%%%%%%%%%%%%%%%%%%%%%%%%%%%%%%%%%%%%%%%%%%%%%%%%%%%%
% ACKNOWLEDGEMENTS
%%%%%%%%%%%%%%%%%%%%%%%%%%%%%%%%%%%%%%%%%%%%%%%%%%%%%%%%%%%%%%%%%%%%%%%%%%%%%%%%%%%%%%%%%%%%%%%%%%%%%%%%%%%%%%%%%%%%%
%\vspace{0.5cm}
\subsection*{Acknowledgements}

I warmly thank Gianguido Dall'Agata, Antonio Masiero, Marco Matone, Massimo Pietroni and Roberto Volpato for useful conversations.
%%%%%%%%%%%%%%%%%%%%%%%%%%%%%%%%%%%%%%%%%%%%%%%%%%%%%%%%%%%%%%

%%%%%%%%%%%%%%%%%%%%%%%%%%%%%%%%%%%%%%%%%%%%%%%%%%%%%%%%%%%%%%%%%%%%%%%%%%%%%%%%%%%%%%%%%%%%%%%%%%%%%%%%%%%%%%%%%%%%%
% BIBLIOGRAPHY
%%%%%%%%%%%%%%%%%%%%%%%%%%%%%%%%%%%%%%%%%%%%%%%%%%%%%%%%%%%%%%%%%%%%%%%%%%%%%%%%%%%%%%%%%%%%%%%%%%%%%%%%%%%%%%%%%%%%%

\providecommand{\href}[2]{#2}\begingroup\raggedright\endgroup

\end{document}